\documentclass[trackchanges]{aastex701}

\begin{document}

\title{Galaxies caught in transition: the role of group environment in shaping the mass–size relation in the local Universe}

\correspondingauthor{Gissel P. Montaguth}
\email{gissel.pmontaguth@usp.br}

\author[orcid=00009-0003-1364-3590,gname=Gissel, sname=P. Montaguth]{Gissel P. Montaguth}
\affiliation{Instituto de Astronomia, Geof\'isica e Ci\^encias Atmosf\'ericas da Universidade de S\~ao Paulo, Cidade Universit\'ria, CEP:05508-990, S\~ao Paulo, SP, Brazil}
\email{gissel.pmontaguth@usp.br}

\author[orcid=0000-0002-5267-9065,gname=Claudia Laura, sname=Mendes de Oliveira]{Claudia Mendes de Oliveira}
\affiliation{Instituto de Astronomia, Geof\'isica e Ci\^encias Atmosf\'ericas da Universidade de S\~ao Paulo, Cidade Universit\'ria, CEP:05508-990, S\~ao Paulo, SP, Brazil}
\email{claudia.oliveira@iag.usp.br}

\author[orcid=0009-0006-0373-8168,gname=Ciria, sname=Lima-Dias]{Ciria Lima-Dias}
\affiliation{Departamento de Astronomía, Universidad de La Serena, Avda. Ra\'ul Bitr\'an 1305, La Serena, Chile}
\email{clima@userena.cl}

\author[orcid=0000-0003-2325-9616,gname=Antonela,sname=Monachesi]{Antonela Monachesi}
\affiliation{Departamento de Astronomía, Universidad de La Serena, Avda. Ra\'ul Bitr\'an 1305, La Serena, Chile}
\email{amonachesi@userena.cl}

\author[orcid=0000-0002-7005-8983,gname=Sergio,sname=Torres-Flores]{Sergio Torres-Flores}
\affiliation{Departamento de Astronomía, Universidad de La Serena, Avda. Ra\'ul Bitr\'an 1305, La Serena, Chile}
\email{sptorres@userena.cl}

\author[orcid=0000-0002-8280-4445,gname=Eduardo,sname=Telles]{Eduardo Telles}
\affiliation{Observat\'orio Nacional, Rua General Jos\'e Cristino, 77,  S\~ao Crist\'ov\~ao, 20921-400 Rio de Janeiro, RJ, Brazil}
\email{etelles@on.br}

\author[orcid=0000-0001-7907-7884,gname=Fábio,sname=Herpich]{Fábio R. Herpich}
\affiliation{Laboratório Nacional de Astrofísica (LNA/MCTI), Rua Estados Unidos, 154, Itajubá 37504-364, Brazil}
\email{fabiorafaelh@gmail.com}

\author[orcid=0000-0002-6090-2853,gname=Yolanda,sname=Jimenez-Teja]{Yolanda Jim\'enez-Teja}
\affiliation{Observat\'orio Nacional, Rua General Jos\'e Cristino, 77, S\~ao Crist\'ov\~ao, 20921-400 Rio de Janeiro, RJ, Brazil}
\affiliation{Instituto de Astrof\'isica de Andaluc\'ia (IAA-CSIC), Glorieta de la Astronom\'ia s/n, E-18008 Granada, Spain}
\email{yojite@iaa.csic.es}

\author[orcid=0000-0002-2484-7551,gname=Antonio,sname=Kanaan]{Antonio Kanaan}
\affiliation{Departamento de F\'isica, Universidade Federal de Santa Catarina, Florian\'opolis, SC, 88040-900, Brazil}
\email{ankanaan@gmail.com}

\author[orcid=0000-0002-0138-1365,gname=Tiago,sname=Ribeiro]{Tiago Ribeiro}
\affiliation{Rubin Observatory Project Office, 950 N. Cherry Ave, Tucson 85719, USA}
\email{tiago.astro@gmail.com}

\author[orcid=0000-0002-4064-7234,gname=William,sname=Schoenell]{William Schoenell}
\affiliation{GMTO Corporation, N. Halstead Street 465, Suite 250, Pasadena, CA 91107, United States}
\email{wschoenell@gmail.com}

\begin{abstract}
The stellar mass–size relation is a sensitive probe of how environment shapes galaxy structure. We analyse this relation in the local Universe for galaxies in compact groups (CGs), low-mass groups ($M_{vir} \leq 10^{13}~M_{\odot}$), and high-mass groups, comparing them to field galaxies using data from the Southern Photometric Local Universe Survey. Galaxies are classified as early types (ETGs; $n \geq 2.5$, $(u - r)_0 \geq 2.2$), late types (LTGs; $n < 2.5$, $(u - r)_0 < 2.2$), transition galaxies (TGs; $n < 2.5$, $(u - r)_0 \geq 2.2$), and others (OGs; $n \geq 2.5$, $(u - r)_0 < 2.2$). We find that ETGs and OGs show no significant environmental dependence: their mass–size slopes and intercepts are statistically consistent across CGs, groups, and the field. LTGs also follow similar relations in the field and in most groups, with only a modest tendency for LTGs in CGs to be smaller at fixed stellar mass. By contrast, TGs display a clear environmental signal: in groups the slope steepens to $\alpha \sim 0.4$ (versus $\alpha \sim 0.2$ in the field) and their sizes are smaller than in the field, with non-overlapping 95\% posterior intervals. These trends suggest that TGs in denser environments are more structurally evolved, likely owing to enhanced bulge prominence and fading of the outer disc, consistent with the S\'ersic-index distributions, which show an excess of TGs with $n_r \gtrsim 1.5$ in groups and CGs. Our findings highlight TGs as an environmentally sensitive population, providing insight into the structural transformation of galaxies in group environments.

\end{abstract}

\keywords{\uat{Galaxy groups}{597}, \uat{Galaxy interactions}{600}, \uat{Galaxy evolution}{594}, \uat{Galaxy environments}{2029}}

\section{Introduction}

The stellar mass–size relation reflects fundamental aspects of galaxy formation and evolution, shaped by the interplay between internal processes—such as star formation, active galactic nuclei (AGN) activity, and stellar feedback—and external influences, including interactions and mergers \citep{2010Peng, 2019Pallero, 2024Lopes}. Consequently, the stellar mass–size relation serves as a powerful tracer of galaxy assembly \citep{2014VanderWel, 2016Capellari, Suess_2021}. At fixed stellar mass, early-type galaxies (ETGs) are systematically smaller and exhibit steeper mass–size slopes than late-type galaxies (LTGs) \citep{2003Shen, 2014VanderWel, 2015Lange, 2021Nedkova}.


There is a broad consensus that galaxy size is closely linked to mass assembly history and to the properties of their host dark matter halos \citep{1998Mo, 2013Kravtsov}. LTGs typically grow inside-out through star formation and gas accretion \citep{2009Brooks}, whereas ETGs, exhibit a well-established kinematic dichotomy between fast and slow rotators \citep{2011Emsellem,2016Capellari}. Fast rotator ETGs share many structural properties with discs and are thought to evolve largely from spiral progenitors through a combination of gas accretion, secular evolution and relatively minor mergers, often retaining significant ordered rotation \citep{2011Khochfar}. In contrast, the most massive slow rotators are generally associated with a two-phase formation scenario: at early times ($z \gtrsim 2-3$) they build compact, dense cores through highly dissipative processes and gas-rich major mergers \citep{2025DeGraaff, 2025Ito}, while at later epochs ($z \lesssim 1-2$) their subsequent size growth and the build-up of extended envelopes are dominated by dry mergers, in particular multiple minor mergers that add ex-situ stars to their outskirts and reduce their specific angular momentum \citep{2009Naab, 2010Oser}.

Within this overall framework, \citet{2016Capellari} further argues that the transformation from LTGs to ETGs follows two main evolutionary paths: one driven by gas accretion and bulge growth, producing passive fast-rotators with moderate size increase, and another dominated by dry mergers, leading to slow-rotators with sizes growing roughly with stellar mass. They propose that the mass–size relation is largely universal across environments, with environmental effects emerging mainly through shifts in the morphological mix—spirals being replaced by fast-rotator ETGs in denser regions. However, this issue remains debated: some studies report that, at fixed stellar mass, galaxies are generally smaller in dense environments \citep{2013Poggiati, 2014Cebrian, 2023Montaguth}, while others find no significant environmental dependence \citep{Rettura_2010, 2013Huertas-Companyb}. These conflicting results are likely driven by differences in sample selection, definitions of environment, and the methods used to measure galaxy properties. This ambiguity underscores the need for a differential analysis that can isolate galaxy populations still actively responding to their environment, rather than those whose structures have already been set by their integrated mass assembly history.

Further evidence points to environmental trends: \citet{2025Perez} found that ETGs in voids are 10–20\% smaller than in denser environments, likely due to reduced merger activity. For LTGs, environmental effects appear mainly at low stellar masses, where they tend to be smaller in clusters, likely as a consequence of gas stripping and tidal interactions. Using bulge–disk decompositions, \citet{2017Kuchner} showed that cluster-related processes like ram-pressure stripping affect disk outskirts, leading to more compact disks. In Hydra, \citet{2024Lima-Dias} found that disk size increases with stellar mass, while bulge size remains largely uncorrelated, suggesting that disks dominate the mass–size relation in dense environments—unlike in the field, where both components scale with stellar mass \citep{2021Mendez}.

In this context, galaxy groups—where interactions are typically more prolonged—represent a critical laboratory for understanding galaxy evolution. These environments are thought to play a central role in shutting down star formation and driving morphological transformations, acting as key sites of pre-processing before galaxies fall into massive clusters \citep{1998Zabludoff,2005Wilman,2004Eke,2012Coenda,2023Montaguth,2024Oxland, 2024Lopes,2024Epinat}. Yet, while the stellar mass–size relation has been extensively studied in clusters, its behavior in galaxy groups has received comparatively less attention, despite the fact that at least half of the galaxies in the Universe reside in such environments \citep{2004Eke}. Our study directly addresses this gap by investigating how the group environment influences the stellar mass–size relation in the local Universe ($0.03 < z < 0.1$). We focus on compact groups, a small system of typically between three and ten galaxies in close proximity on the sky, with a characteristic radius of a few tens of kiloparsecs \citep[e.g.][]{hickson1982systematic,zheng2020compact}, low-mass groups ($M_{\rm vir} \leq 10^{13}~M_\odot$), and high-mass groups ($M_{\rm vir} > 10^{13}~M_\odot$), comparing them with a control sample of field galaxies.

The paper is organized as follows: in Section~\ref{sec:data-method}, we describe the data, sample selection, and methodology; in Section~\ref{sec:results}, we present the mass–size relation results for each morphological type; in Section~\ref{sec:discussion}, we discuss the physical implications of our findings; and in Section~\ref{sec:summary}, we summarize our main conclusions. Throughout the paper, we adopt a flat cosmological model with parameters $H_0 = 70 km$ $s^{-1}$ $Mpc^{-1}$, $\Omega_M = 0.3$, and $\Omega_\lambda = 0.7$ (\citealt{2003Spergel}).

\section{Data and Methodology}
\label{sec:data-method}

To explore the environmental dependence of the mass–size relation, we combine galaxies in groups drawn from the 2dFGRS Percolation-Inferred Galaxy Group (2PIGG) catalog \citep{2004Eke}, compact groups (CGs), and a field sample from \citet{2023Montaguth}. All samples are located within the Southern Photometric Local Universe Survey(S-PLUS) DR4 footprint \citep{2024DR4splus}.S-PLUS is a 12-band optical imaging survey of the southern sky, carried out with the T80-South telescope at CTIO, covering a field of view of $\sim2\mathrm{deg}^2$\citep{mendes2019southern}. It combines five broad-band filters ($u,g,r,i,z$) with seven narrow-band filters (J0378, J0395, J0410, J0430, J0515, J0660, J0861) centred on key spectral features (e.g. [O,II], Ca H+K, H$\delta$, the G band, Mg$b$, H$\alpha$, Ca,II triplet), providing homogeneous, multi-band imaging that is well suited to constraining structural parameters.

We identified CGs from the catalogue by \citet{zheng2020compact}, constructed using SDSS-DR14, LAMOST and GAMA data, where CGs are selected by combining the Hickson  photometric criterion \citep{hickson1982systematic} with spectroscopic information on radial velocities. In this catalogue, a CG must have 3–10 members in the magnitude range $14.00\leq r\leq17.77$, satisfy an isolation condition $\theta_N\geq \theta_G$, where $\theta_G$ is the group radius and $\theta_N$ the distance to the nearest bright galaxy, be compact with  $\mu_r \le 26.0~\mathrm{mag}\,\mathrm{arcsec}^{-2}$, and each member galaxy must have a line of sight velocity difference $|V - V_i| \le 1000~\mathrm{km\,s^{-1}}$ relative to the group. Field galaxies were selected from the \citet{2007yang} catalog as isolated objects, within the same redshift range and, with a similar absolute magnitude distribution as that of the sample of CGs. The general sample of groups was drawn from the 2PIGG catalog, using a friends-of-friends algorithm (\citealt{1985FoF}) applied to the 2dFGRS data (\citealt{20012dFGRS}), requiring a minimum of three members. To ensure a fair comparison across environments, the sample of groups was restricted to match the absolute magnitude ($M_r$) and redshift distributions of the CG and field galaxy samples.

\begin{figure}
    \centering
    \includegraphics[width=0.45\textwidth]{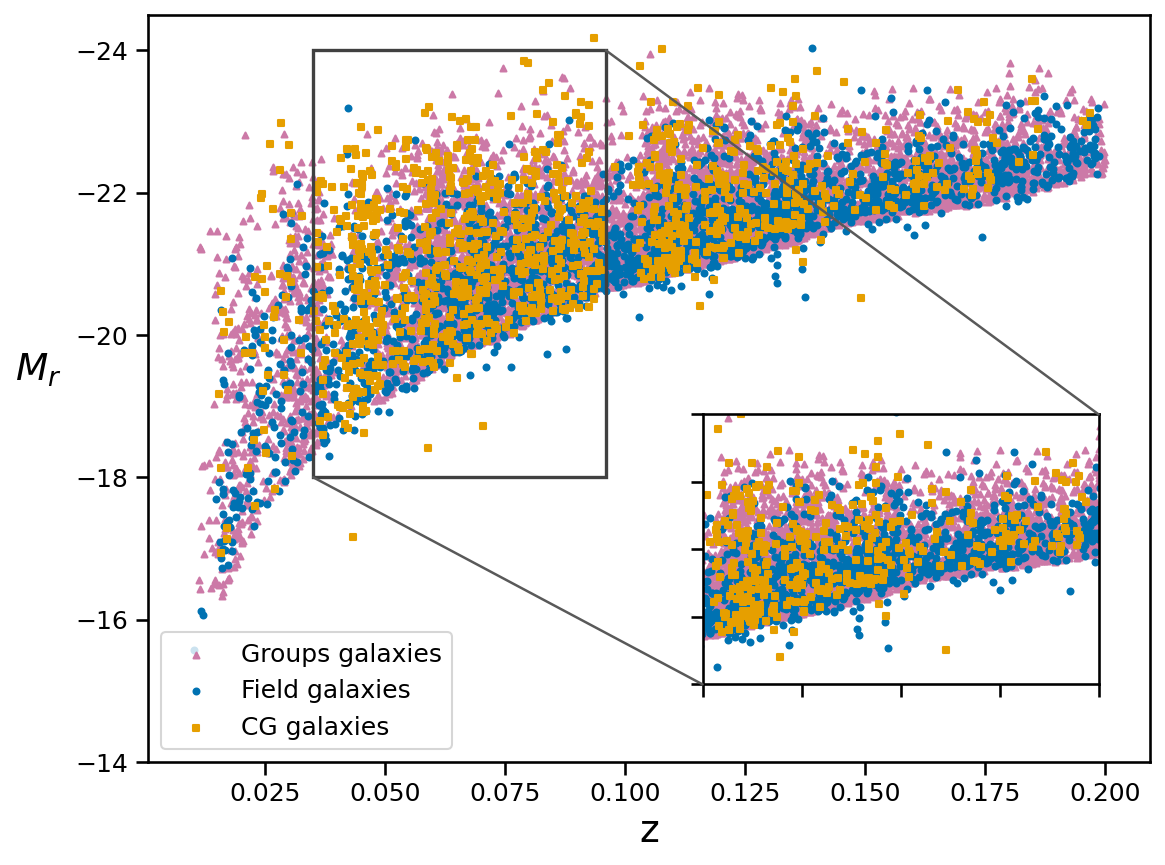}
    \caption{Distribution of absolute $r$-band magnitude ($M_r$) as a function of redshift ($z$) for the parent galaxy sample. The gray rectangle marks the redshift interval adopted in this work, $0.035 \le z < 0.096$. The inset shows a magnified view of this region, which defines the final sample analyzed. Points are color-coded by environment: group galaxies (purple), field galaxies (blue), and compact-group (CG) galaxies (orange).}
    \label{fig:mr-z}
\end{figure}

This approach ensures consistency in stellar mass estimates and reduces selection biases in the analysis of the mass–size relation. The sample consists of 10385 galaxies: 7021 in groups, 1083 in CGs, and 2281 in the field. We then select a redshift range of $0.035 < z < 0.096$ to ensure greater homogeneity among the three samples. Below this redshift range, the CG population is under-represented compared to the field and group samples, while above $z \sim 0.096$ there is a gap in the CG sample coverage. After applying this redshift cut, the final sample comprises 2402 group galaxies, 595 CG galaxies, and 915 field galaxies (used as the control sample). Figure~\ref{fig:mr-z} shows the distribution of absolute magnitude versus redshift for the final sample.

We derive the structural parameters using the \textsc{MorphoPLUS} pipeline \citep{Montaguth2025b}, which performs Sérsic profile \citep{Sersic} fitting across all 12 S-PLUS bands using \textsc{GALFITM} \citep{2011Bamford, 2013Haussler, 2013vika}. Therefore, we have access to measurements of effective radius ($R_e$) and Sérsic index ($n$) in twelve optical filters. However, for the purposes of this study, we focus exclusively on the $r$ band. Our analysis may be affected by observational and selection biases. Sérsic profile fits are sensitive to image quality, and both the seeing, which varies between $\sim 0.82\arcsec$ and $2\arcsec$ \citep{mendes2019southern}, and the survey depth, which for the main survey corresponds to a S/N $\sim$10 Petrosian limit of $r \simeq 19.5$~mag (and $u \simeq 19.3$~mag; see Table~2 of \citealt{2024DR4splus}), can introduce biases in the derived galaxy parameters. Nevertheless, direct comparisons with DECaLS \citep{OrtizGomez2025} reveal only minor systematic offsets ($\Delta n \sim 0.1$, $\Delta R_e \sim 0.5$ kpc), well within the measurement uncertainties. The catalogs used also have limitations (e.g. interlopers in 2PIGG, projection effects in CGs), but a full quantification of these effects is beyond the scope of this work.

We estimate the stellar masses using the $K$-corrected rest-frame $(g - i)_0$ color and the absolute $i$-band magnitude, following the method by \citet{2011Taylor}, which assumes a \citet{2003Chabrier} initial mass function, the \citet{2003Bruzual} stellar population synthesis models, and the \citet{2001Calzetti} dust law, and provides a simple and well-tested proxy for optical SED fitting. As discussed by \citet{2011Taylor}, at fixed $(g-i)_0$ the intrinsic scatter in $M_{*}/L_i$ is typically $\sim 0.1$~dex, whereas the global systematics associated with SPS based stellar masses (e.g. choice of models, IMF, and dust law) are expected to be of order $0.2$–$0.3$~dex \citep[e.g.][]{2009Conroy,Salim2016}. In \citet{2023Montaguth} we explicitly quantified the differences between stellar masses derived from the \citet{2011Taylor} colour–based calibration using S-PLUS data and SED–based masses from \citet{Salim2016}, finding a typical scatter of 0.13~dex and a systematic offset that increases to $\sim 0.2$~dex at the highest stellar masses in our sample ($M_{*} \gtrsim 10^{10}\,{\rm M_\odot}$). These values lie comfortably within the $\sim 0.2$–$0.3$~dex systematic uncertainties expected between independent SPS based stellar mass estimators.

\begin{figure}
    \centering
    
    \includegraphics[width=0.45\textwidth]{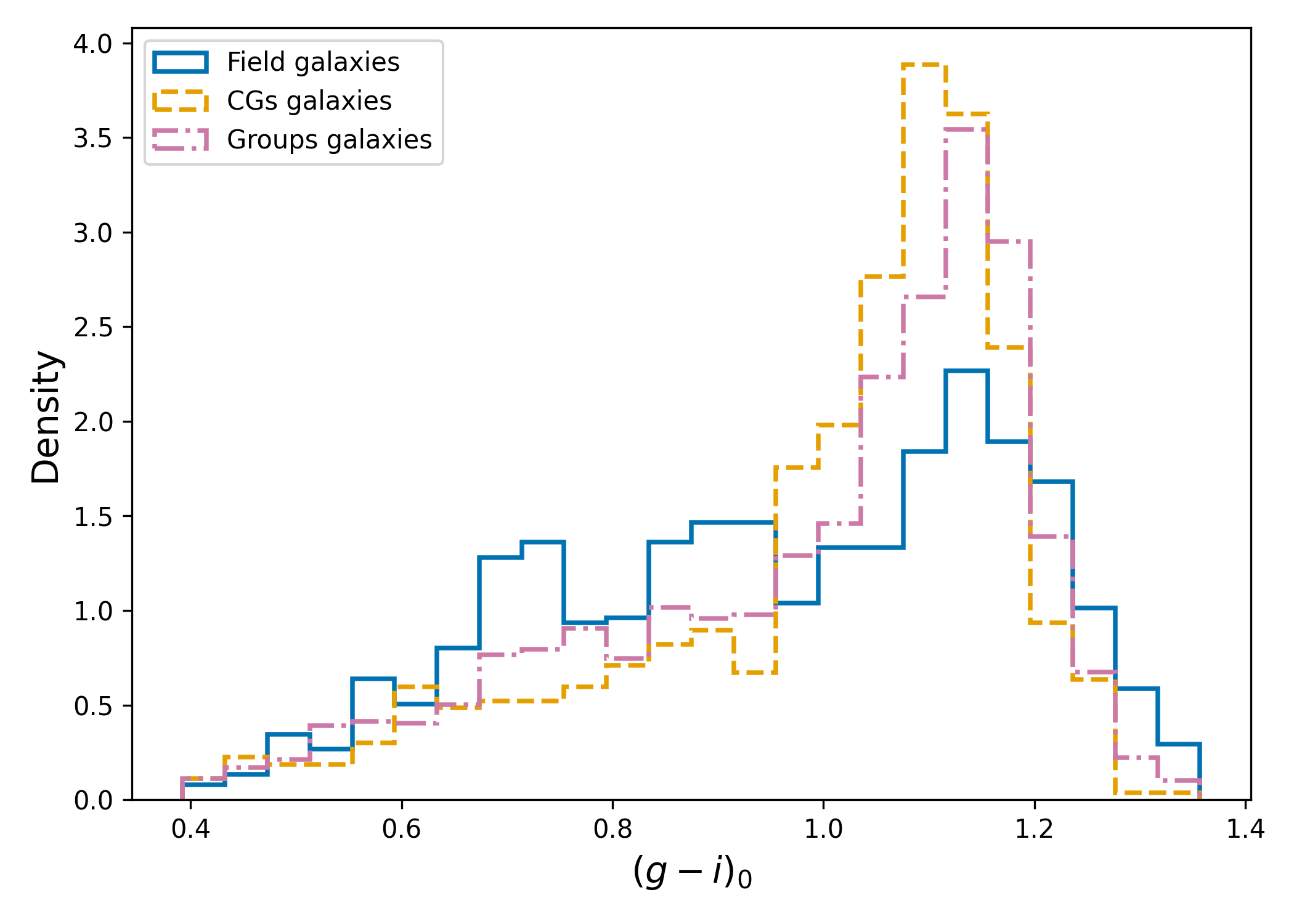}
    \includegraphics[width=0.45\textwidth]{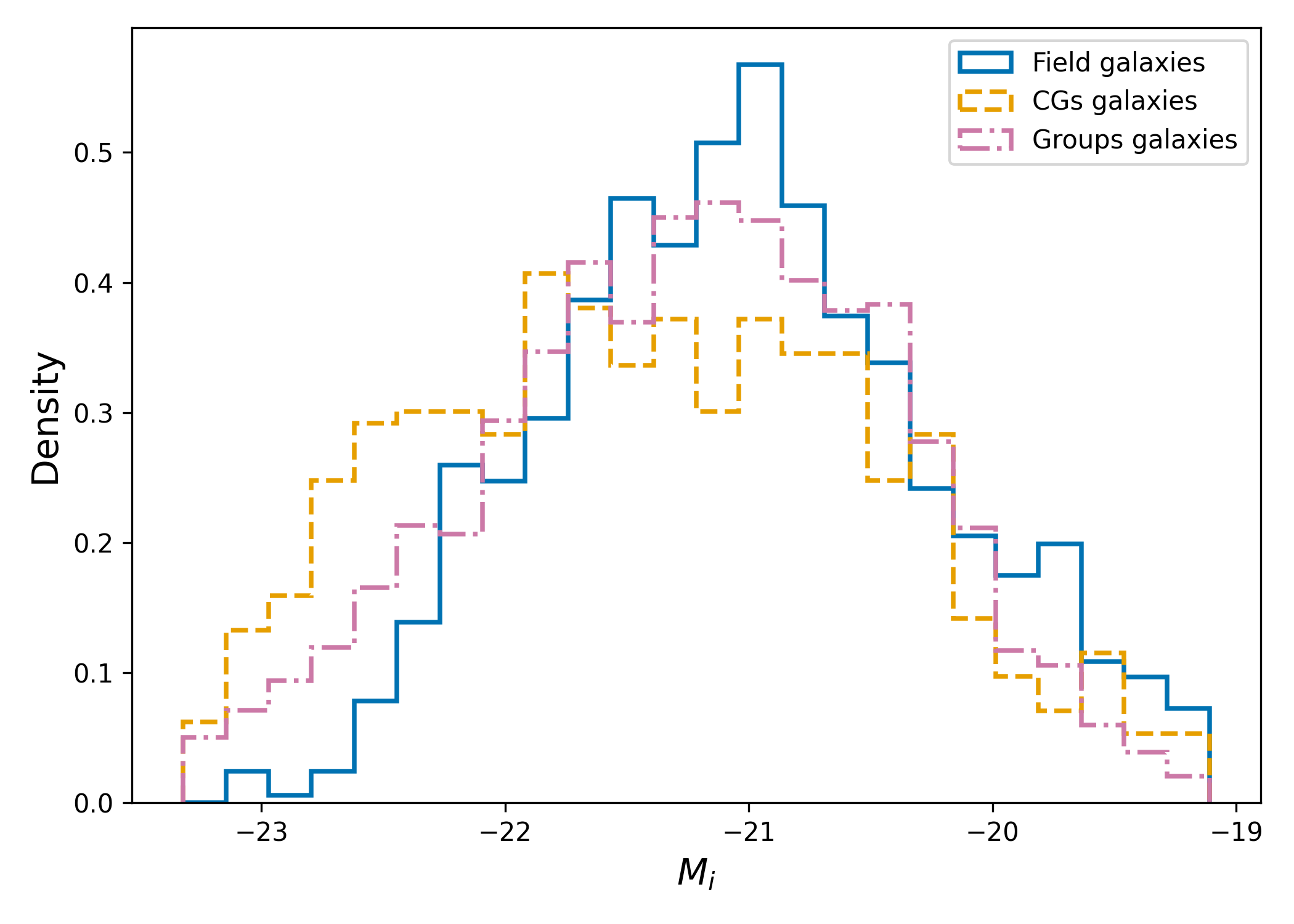}
    \includegraphics[width=0.45\textwidth]{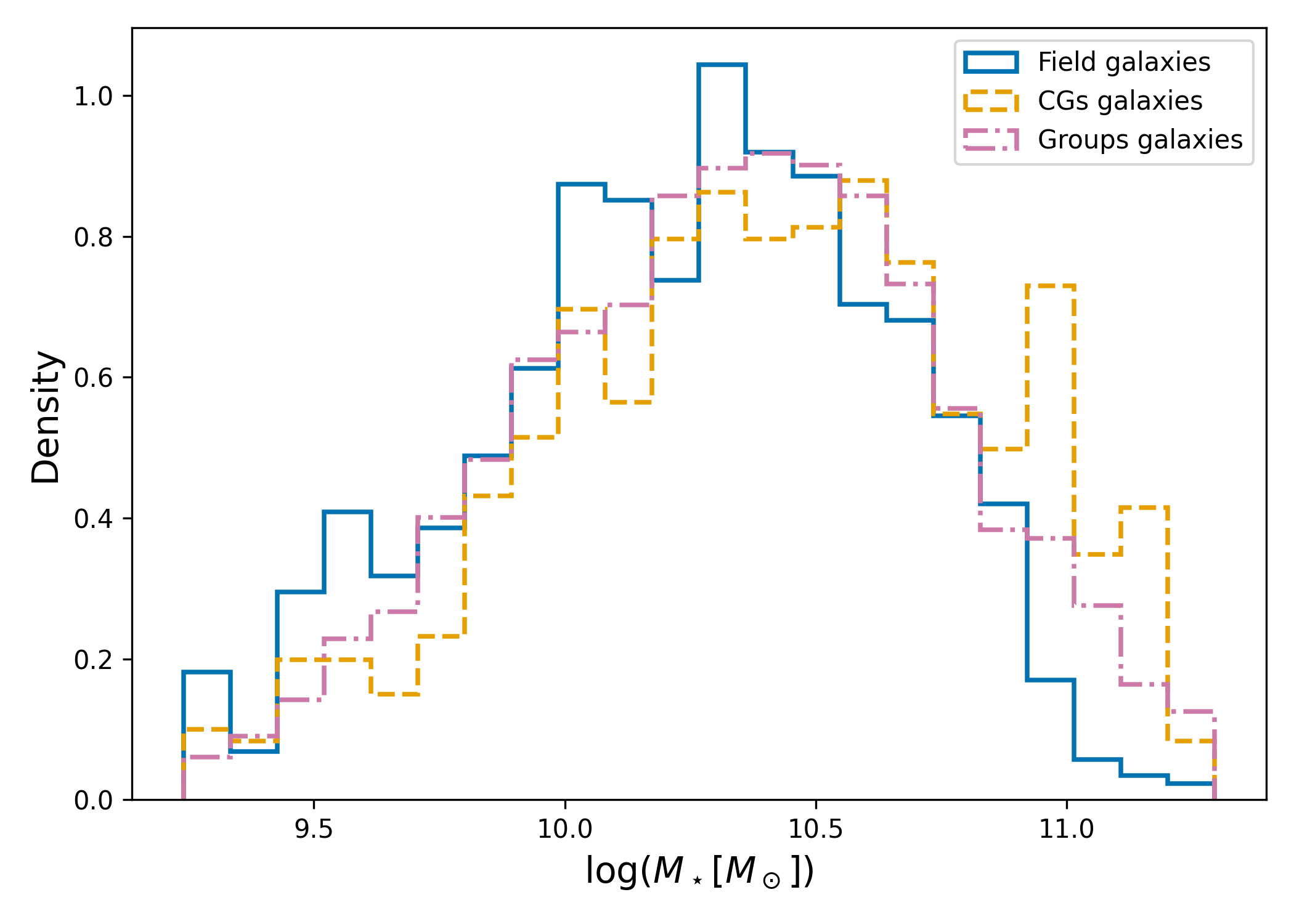}
    \caption{
    Density distributions of rest–frame $(g-i)_0$ colour (top-left panel), absolute $i$-band magnitude $M_i$ (top-right panel), and stellar masses (bottom panel) for field galaxies (blue), CGs members (orange), and galaxies in groups (green).}
    \label{fig:col_mag_dists}
\end{figure}

Because our stellar masses are inferred from optical colour and $i$-band luminosity, it is important to verify that the different environmental samples are broadly comparable in both quantities. Figure~\ref{fig:col_mag_dists} shows the denstity distributions of rest–frame $(g-i)_0$, absolute $i$-band magnitude $M_i$, and the stellar masses ($M_*$) for field galaxies (blue histogram), CGs members (orange histogram), and galaxies in the groups sample (purple histogram). All three samples span very similar ranges in $M_*$, $(g-i)_0$ and $M_i$, and their overall shapes are broadly comparable. CGs and group galaxies display an excess of red systems with $(g-i)_0 \gtrsim 1$ and modest excess at the high-mass end ($\log(M*[M_{\odot}]) \geq 10.7-11$), consistent with the well-known relation between environment and galaxy colour \citep{2006Baldry,2010Peng}, with denser regions hosting a higher fraction of red, and quenched galaxies. In $M_i$, the CG sample is slightly biased toward brighter magnitudes. This is a natural consequence of the CGs selection, which favours systems with relatively bright members. Despite these trends, there is substantial overlap among the three samples in colour and luminosity, giving us confidence that our environmental comparisons are not driven by gross mismatches in the underlying stellar mass or colour distributions.

As an additional consistency check to ensure that the small differences in $M_*$ reflect environmental effects rather than a correlation between the $(u-r)$ and $(g-i)$ colours within each morphological class, we compared our colour-based stellar masses with SED-derived stellar masses from \cite{Salim2016} for the field and CG samples, which are the subsamples with \cite{Salim2016} data available (see also \cite{2023Montaguth}). The resulting ($\Delta \log M \equiv \log(M_{*,\mathrm{SED}}) - \log(M_{*,\mathrm{colour}})$) distributions are statistically consistent between environments: two-sample 
Kolmogorov-Smirnov (KS) tests yield p-values in the range $0.051–0.586$ across the morphological classes, and the robust scatter is comparable in both samples ($\sigma_{\mathrm{MAD}} \approx 0.06–0.13$ dex).

\begin{table}[ht]
\centering

\caption{Morphological type distribution in different environments. The error is the counting uncertainty associated with the number of events, calculated under the assumption of Poisson statistics.}
\label{tab:morpho}
\begin{tabular}{lccc}
\hline
\textbf{Type} & \textbf{Field (\% ± err)} & \textbf{Group (\% ± err)} & \textbf{CG (\% ± err)} \\
\hline
ETG & $21.4 \pm 1.0$ & $38.8 \pm 0.6$ & $49.7 \pm 1.5$ \\
LTG & $49.6 \pm 1.0$ & $36.0 \pm 0.5$ & $29.9 \pm 1.3$ \\
TG  & $17.4 \pm 0.8$ & $10.6 \pm 0.3$  & $11.6 \pm 0.9$  \\
OG  & $11.6 \pm 0.7$ & $14.6 \pm 0.4$ & $8.8 \pm 1.1$ \\
\hline
\end{tabular}
\end{table}

\begin{figure}
    \centering
    \includegraphics[scale=0.5]{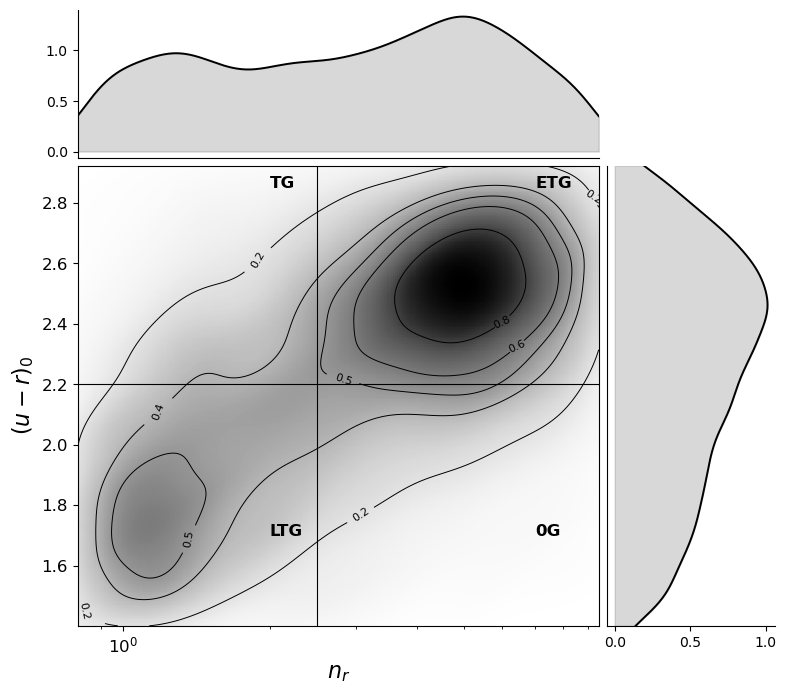}
    \caption{Classification of ETGs, transition galaxies, and LTGs based on rest–frame colour $(u-r)_0$ and the Sérsic index in the r-band ($n_r$). The vertical guide marks ($n_r=2.5$), and the horizontal guide marks $(u-r)_0=2.2$. A greyscale background displays the two–dimensional kernel–density estimate of the joint distribution, and the marginal normalised histograms are shown along the top $(u-r)_0$ axis and the right $log(n_r)$ axis.}
   \label{fig:vika}
\end{figure}

We classify galaxies based on their Sérsic index $n_r$ in the $r$ band and $K$-corrected rest-frame $(u - r)_0$ colour  \citet{Chilingarian2010}, corrected for Galactic extinction by using \cite{1989Cardelli}, extinction law with $R_v=3.1$, and the maps from \cite{2011schlafly}. Our scheme is inspired by the approach of \citet{vika2015megamorph}, who used fixed thresholds of $(u-r)=2.3$ and $n_r=2.5$ to separate blue abd disc-dominated systems from red and spheroid-dominated ones. However, we adopt the data-driven boundaries introduced in \citet{Montaguth2025b}. In that work we combined group galaxies with a homogeneous control sample of field galaxies and showed that the $(u-r)_0$–$\log(n_r)$ plane is strongly bimodal. Using two-component Gaussian Mixture Models fitted to the unbinned one-dimensional distributions of $(u-r)_0$ and $\log(n_r)$, \citet{Montaguth2025b} identified the intersections of the components at $(u-r)_0 \simeq 2.22$ and $n_r \simeq 2.54$, which we round to $(u-r)_0 = 2.2$ and $n_r = 2.5$ and adopt here as objective, data-driven boundaries between the blue/low-$n_r$ and red/high-$n_r$ populations. Figure \ref{fig:vika} shows the $(u-r)_0-n_r$ distribution for the combined field, group, and CG samples in this work. The grayscale kernel–density map, together with its iso-density contours and the marginal distributions, confirms the clear bimodality in this plane. The lines at $(u-r)_0=2.2$ and $n_r = 2.5$ follow the low–density valley between the two main concentrations, consistent with their use as boundaries between the main morphological classes.

Early-type galaxies (ETGs) are defined as systems with $n_r \geq 2.5$ and $(u - r)_0 \geq 2.2$; late-type galaxies (LTGs) as those with $n_r < 2.5$ and $(u - r)_0 < 2.2$; transition galaxies (TGs) as systems with $n_r < 2.5$ and $(u - r)_0 \geq 2.2$; and other galaxies (OGs) as those with $n_r \geq 2.5$ and $(u - r)_0 < 2.2$. Table~\ref{tab:morpho} summaries the distribution of morphological types for each environment. In addition, we divide the group sample into two categories: low-mass groups ($M_{\rm vir} \leq 10^{13}\,M_\odot$) and high-mass groups ($M_{\rm vir} > 10^{13}\,M_\odot$). The virial mass $M_{\rm vir}$, estimated following the prescription described in \citet{2024Epinat}, is derived from the line-of-sight velocity dispersion of group members, estimated with the gapper method \citep{1990Beers}. This dispersion allows us to determine $R_{200}$ and, under the assumption of dynamical equilibrium, the virial radius and virial mass.

An important point to emphasise is that our results should be interpreted in the light of a hybrid structural–colour classification, which is calibrated against sSFR in \citet{2023Montaguth}. In that work, ETGs and LTGs correspond, respectively, to predominantly quenched, high-$n_r$ systems and actively star-forming, low-$n_r$ discs, while TGs and OGs trace intermediate regimes in which structure and star-formation state are decoupled. TGs are, on average, redder and include a higher fraction of quenched, S0-like discs with disturbed morphologies, whereas OGs comprise a more diverse population of blue, disc-dominated galaxies, including some early-type systems with residual star formation, which lie outside the canonical LTG region because of their central light concentration. In this sense, both TGs and OGs can be viewed as broadly “transition” populations, analogous to the star-forming ETGs and quenched LTGs identified in studies that treat morphology and star formation separately \citep[e.g.][]{2015Kartaltepe,2023Mei}. In particular, \citet{2015Kartaltepe} showed that visual classifications correlate well with Sérsic indices measured with \textsc{galfit}: galaxies classified as discs tend to have lower Sérsic indices, while those classified as spheroids tend to have higher Sérsic indices, which supports our hybrid structural–colour scheme. In our framework, quenched disc galaxies are naturally mapped into the TG, whereas early-type systems with residual star formation fall into the OG.

Using the stellar mass and effective radius measurements, we model the mass–size relation via a robust Bayesian linear regression \citep{gelman1995bayesian}, adopting a Student-t likelihood for the residuals in order to reduce the influence of outliers and non-Gaussian tails. For each morphological type and environment we model the stellar mass-size relation as:
\begin{equation}
    \log R_{\mathrm{e}} = \alpha + \beta\,\bigl(\log M_\star - 10.5\bigr),
\end{equation}
where the intercept $\alpha$ represents the typical $\log R_{\mathrm{e}}$ at the pivot mass
$\log(M_\star/M_\odot) = 10.5$, and the slope $\beta$ describes the change of size with
stellar mass. The intrinsic scatter at fixed mass is described by a
scale parameter $\sigma$, assuming that $\log R_{\mathrm{e}}$ at given $M_\star$ is
distributed as:
\begin{equation}
    \log R_{\mathrm{e}} \sim \mathrm{Student}\text{-}t\!\left(\nu,\,
    \alpha + \beta\,(\log M_\star - 10.5),\,\sigma\right),
\end{equation}

We sample the posterior distributions of the slope $\beta$, intercept $\alpha$, and
scatter $\sigma$ using Markov Chain Monte Carlo as implemented in \textsc{PyMC}, and report posterior medians and 95\% posterior intervals for all parameters.

\section{Results}
\label{sec:results}

\begin{figure}
    \centering
    \includegraphics[width=0.45\textwidth]{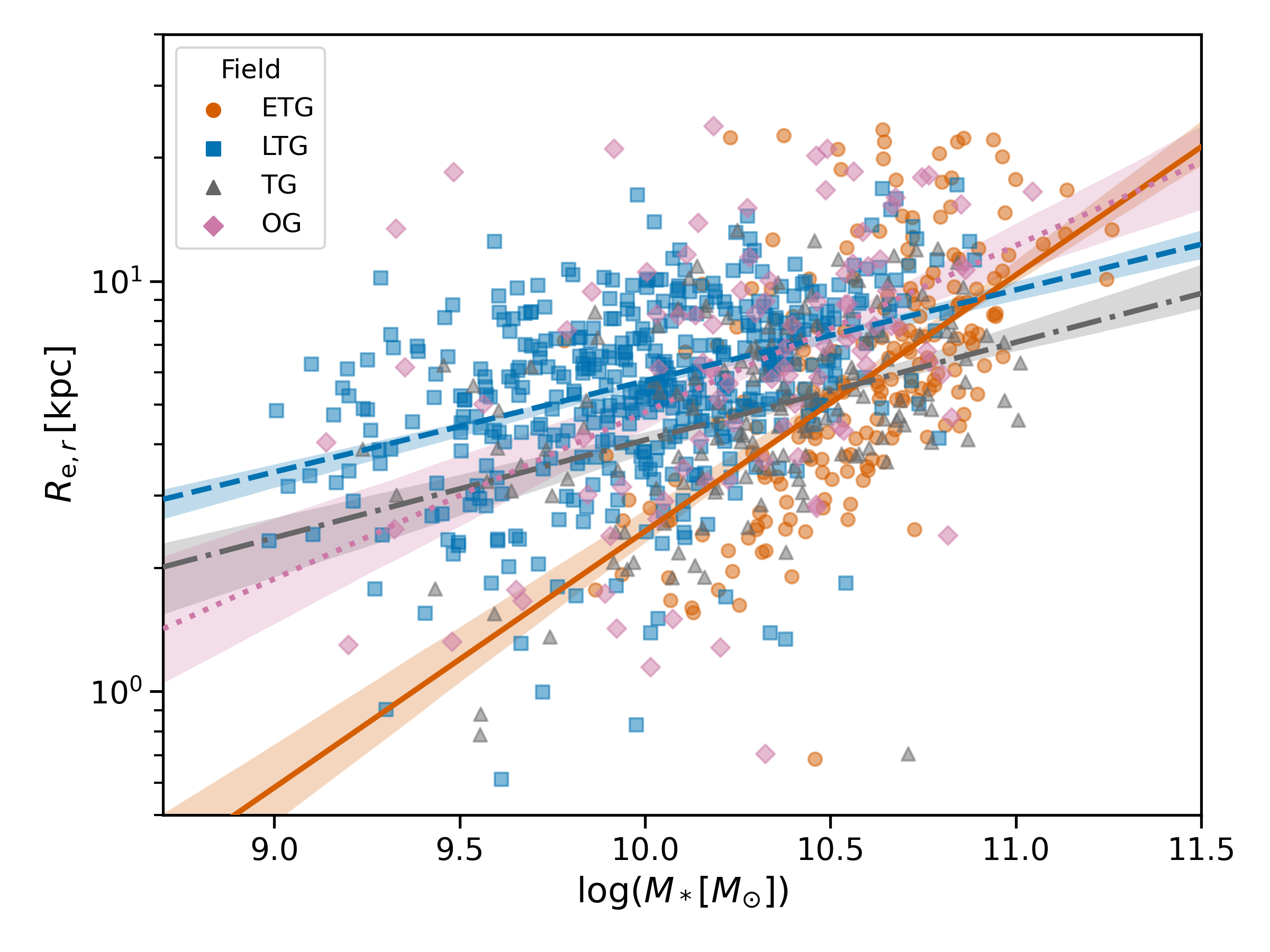}
    \includegraphics[width=0.45\textwidth]{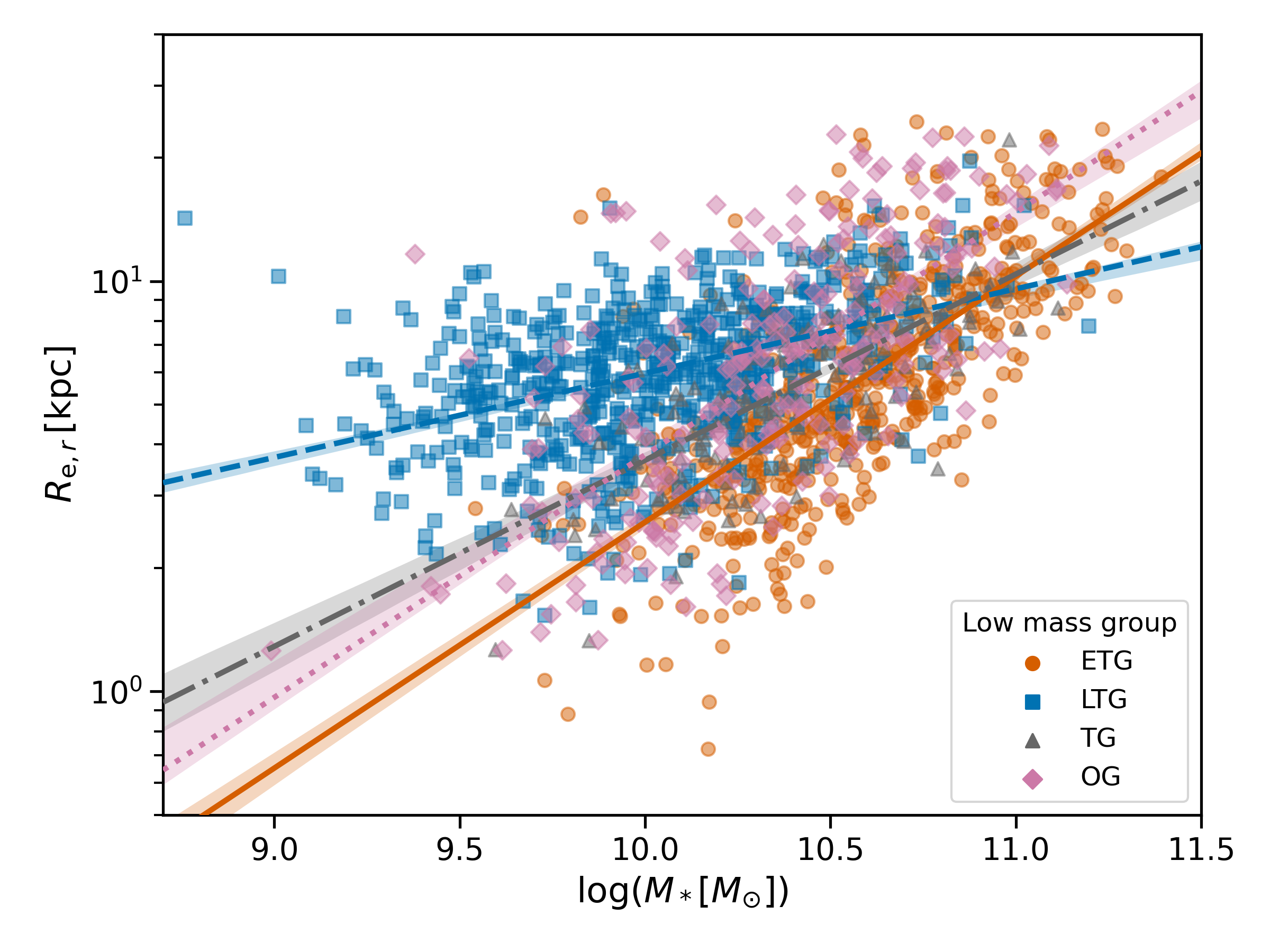}
    \includegraphics[width=0.45\textwidth]{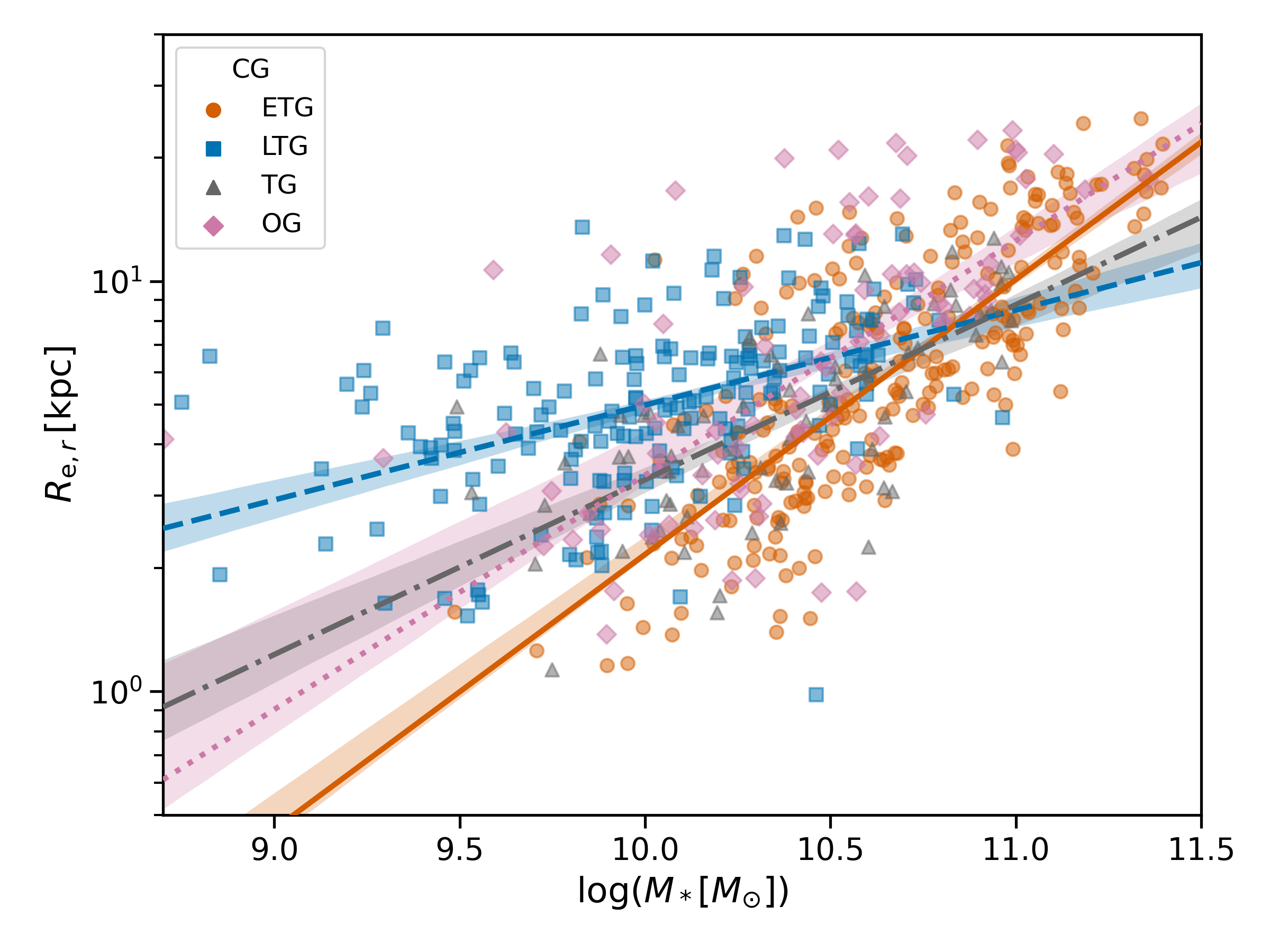}
    \includegraphics[width=0.45\textwidth]{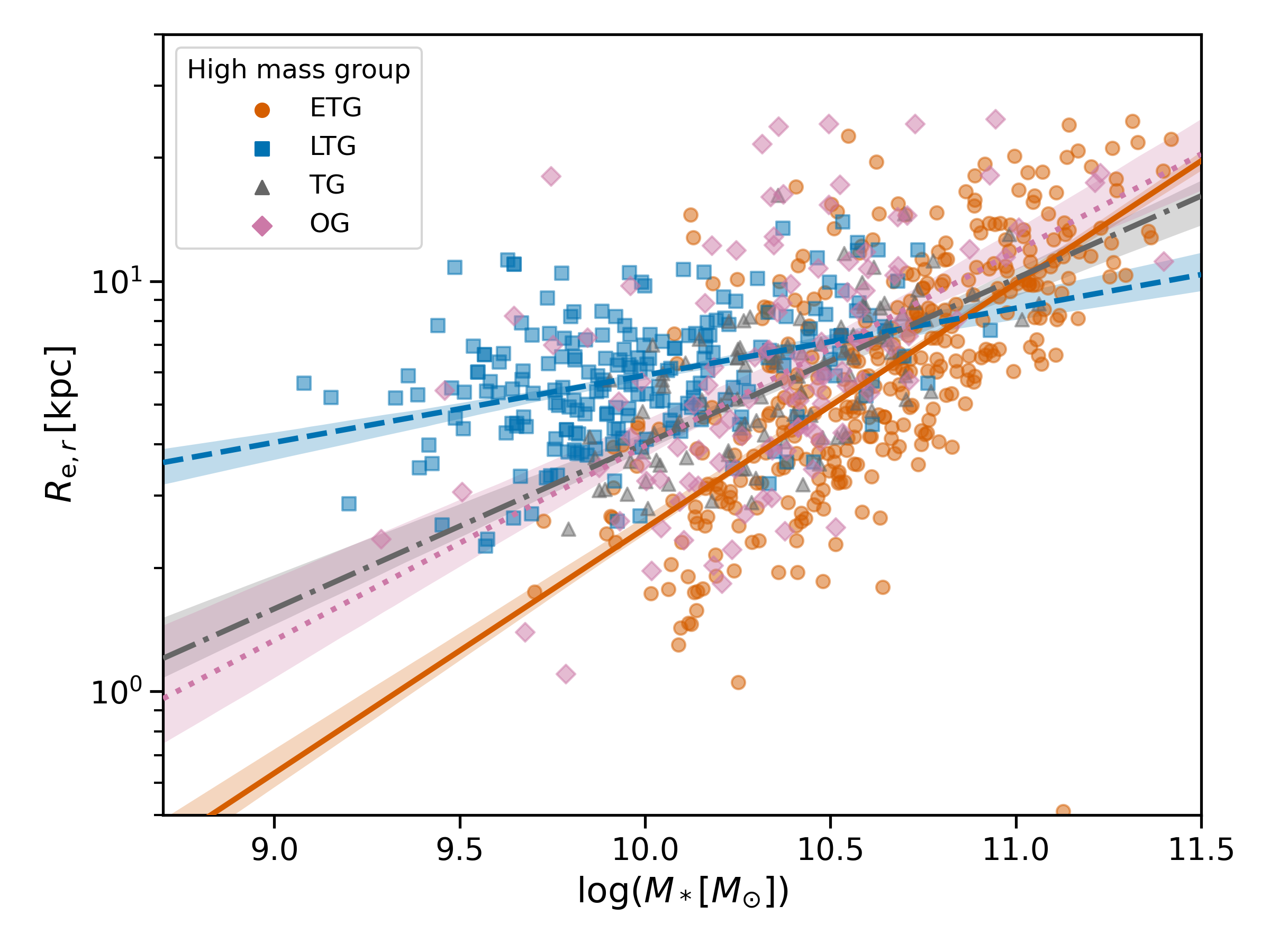}
    
    \caption{Relation between effective radius and stellar mass for galaxies in different environments: field (top-left panel), low-mass groups (top-right panel), compact groups (bottom-left panel), and high-mass groups (bottom-right panel). The colour and symbol of the points indicate the morphological type: orange circles for ETGs, blue squares for LTGs, grey triangles for TGs, and purple diamonds for OGs. For each type, the median mass–size relation is shown as a line (solid for ETGs, dashed for LTGs, dash–dotted for TGs, and dotted for OGs), with shaded regions indicating the bootstrap-estimated uncertainties.}
    \label{fig:re-mass}
\end{figure}

In Figure~\ref{fig:re-mass} we show the relation between effective radius in kiloparsecs, measured in the $r$-band, as a function of stellar mass for each environment: field galaxies (top-left panel), low-mass groups (top-right panel), CGs (bottom-left panel), and high-mass groups (bottom-right panel). The colour and symbol of the points indicate the morphological type: orange circles for ETGs, blue squares for LTGs, grey triangles for TGs, and purple diamonds for OGs. For each type, the median mass–size relation is shown as a line (solid for ETGs, dashed for LTGs, dash–dotted for TGs, and dotted for OGs), with the shaded bands indicating the uncertainty on the relation (16th–84th percentile range from the bootstrap realizations).

Table~\ref{tab:size-mass} summaries the posterior medians and 95\% posterior intervals of the slope, intercept, and intrinsic scatter of the stellar mass–size relation for each morphological type and environment. To identify differences across environments for a given morphological type, we exploit the full posterior distributions of the regression parameters. For each morphological type, we draw matched samples from the posteriors of the field and of each of the other environments (low-mass groups, high-mass groups, and CGs), and compute the posterior distributions of the differences in slope and intercept, $\Delta\beta$ and $\Delta\alpha$. We consider the evidence for a difference to be strong when the 95 per cent credible interval of either $\Delta \beta$ or $\Delta \alpha$ excludes zero (equivalently, $p_{tail}<0.005$). In addition, we quote the posterior probability of the sign of the difference (e.g.  $P(\Delta>0 | data)$) as a quantitative measure of the direction and strength of the evidence. The mass–size slopes we obtain, approximately $\beta \simeq 0.2$ for LTGs and $\beta \simeq 0.5$ for ETGs, are consistent with those reported in the literature \citep{2007Trujillo, 2014VanderWel, 2017Kuchner, 2013Huertas-company}. The inferred intrinsic scatters are modest, with $\sigma_{\mathrm{int}}$ typically around $0.15$ dex in $\log R_{\mathrm{e}}$ and only mildly larger values, up to $\sim 0.24$~dex, for OGs, which is consistent with the expected structural diversity of this class.

\begin{table*}[ht]
\centering

\caption{Mass--size relation parameters for ETGs, LTGs, TGs, and OGs at $0.035 < z < 0.096$. Numbers in brackets indicate the number of galaxies per type in each environment. Columns 2, 4, and 6 list the posterior medians and $1\sigma$ credible intervals of the slope $\beta$, intercept $\alpha$ (at $\log M_\star = 10.5$), and intrinsic scatter $\sigma_{\mathrm{int}}$ in $\log R_{\mathrm{e}}$, respectively. Columns 3 and 5 report two-sided posterior tail-area probabilities for the null hypotheses $\Delta\beta = 0$ and $\Delta\alpha = 0$, computed from the MCMC samples as $p = 2\,\min\!\left[P(\Delta > 0),\,P(\Delta < 0)\right]$. Small values indicate strong evidence for a difference relative to the field.}

\label{tab:size-mass}  
\begin{tabular}{lccccc}
\hline
\textbf{Type – Environment} & \textbf{$\beta$ $\pm$ err} & \textbf{p$_{tail}(\beta)$} & \textbf{$\alpha$ $\pm$ err} & \textbf{p$_{tail}(\alpha)$} & \textbf{$\sigma$ $\pm$ err} \\ \hline

\multicolumn{5}{c}{\textbf{ETGs}} \\ \hline
Field          [209]       & $0.625 \pm 0.054$ & -- & $ 0.703 \pm 0.012$ & -- & $0.171 \pm 0.012$ \\ \hline
Low-Mass Groups  [568]     & $0.600 \pm 0.024$ & $0.667$  & $0.712 \pm 0.077$ & $0.596$ & $0.142 \pm 0.006$  \\ \hline
Compact Groups  [260]      & $0.670 \pm 0.032$  & $0.455$  & $0.669 \pm 0.013$  & $0.092$ & $0.151 \pm 0.009$ \\ \hline
High-Mass Groups  [364]    & $0.590 \pm 0.028$  & $0.637$ & $0.697 \pm 0.011$ & $0.742$ & $0.150 \pm 0.008$ \\ \hline

\multicolumn{5}{c}{\textbf{LTGs}} \\ \hline
Field         [448]        & $0.222 \pm 0.024$ & -- & $0.868 \pm 0.014$ & -- & $0.149 \pm 0.007$\\ \hline
Low-Mass Groups [649]      & $0.206 \pm 0.016$ & $0.590$ & $0.879\pm 0.010$ & $0.519$ & $0.124 \pm 0.005$ \\ \hline
Compact Groups    [178]    & $0.232 \pm 0.033$  & $0.821$ & $ 0.815 \pm 0.019$  & $0.032$& $0.141 \pm 0.010$\\ \hline
High-Mass Groups  [216]    & $0.164\pm 0.028$ & $0.117$ & $0.853\pm 0.017$ & $0.494$ & $0.112 \pm 0.007$  \\ \hline

\multicolumn{5}{c}{\textbf{TGs}} \\ \hline
Field        [159]         & $0.238 \pm 0.046$ & -- & $0.732 \pm 0.017$& --& $0.165 \pm 0.013$ \\ \hline
Low-Mass Groups [154]      & $0.454 \pm 0.038$  & $0.000$ & $0.791 \pm 0.012$ & $0.007$ & $0.116 \pm 0.009$\\ \hline
Compact Groups    [69]    & $0.427 \pm 0.054$ & $0.009$  & $0.729 \pm 0.021$  & $0.920$ & $0.140 \pm 0.016$ \\ \hline
High-Mass Groups [101]     & $0.402 \pm 0.047$ & $0.018$ & $0.813 \pm 0.016$ & $0.001$ & $0.110 \pm 0.010$ \\ \hline

\multicolumn{5}{c}{\textbf{OGs}} \\ \hline
Field    [99]             & $0.404 \pm 0.087$ & -- & $0.883 \pm 0.032$& -- & $0.241 \pm 0.024$ \\ \hline
Low-Mass Groups [237]      & $0.591 \pm 0.042$ & $0.046$ & $0.872 \pm 0.016$  & $0.772$ & $0.186 \pm 0.012$ \\ \hline
Compact Groups    [88]    & $0.572 \pm 0.068$ & $0.127$ & $0.836 \pm 0.024$ & $0.088$ & $0.200 \pm 0.022$ \\ \hline
High-Mass Groups  [113]    & $0.475 \pm 0.068$ & $0.514$  & $0.836 \pm 0.024$ & $0.237$ & $0.195 \pm 0.018$ \\ \hline

\end{tabular}
\end{table*}

\subsection{Comparison between groups and the field}

\begin{figure}
    \centering
    \includegraphics[width=0.46\textwidth]{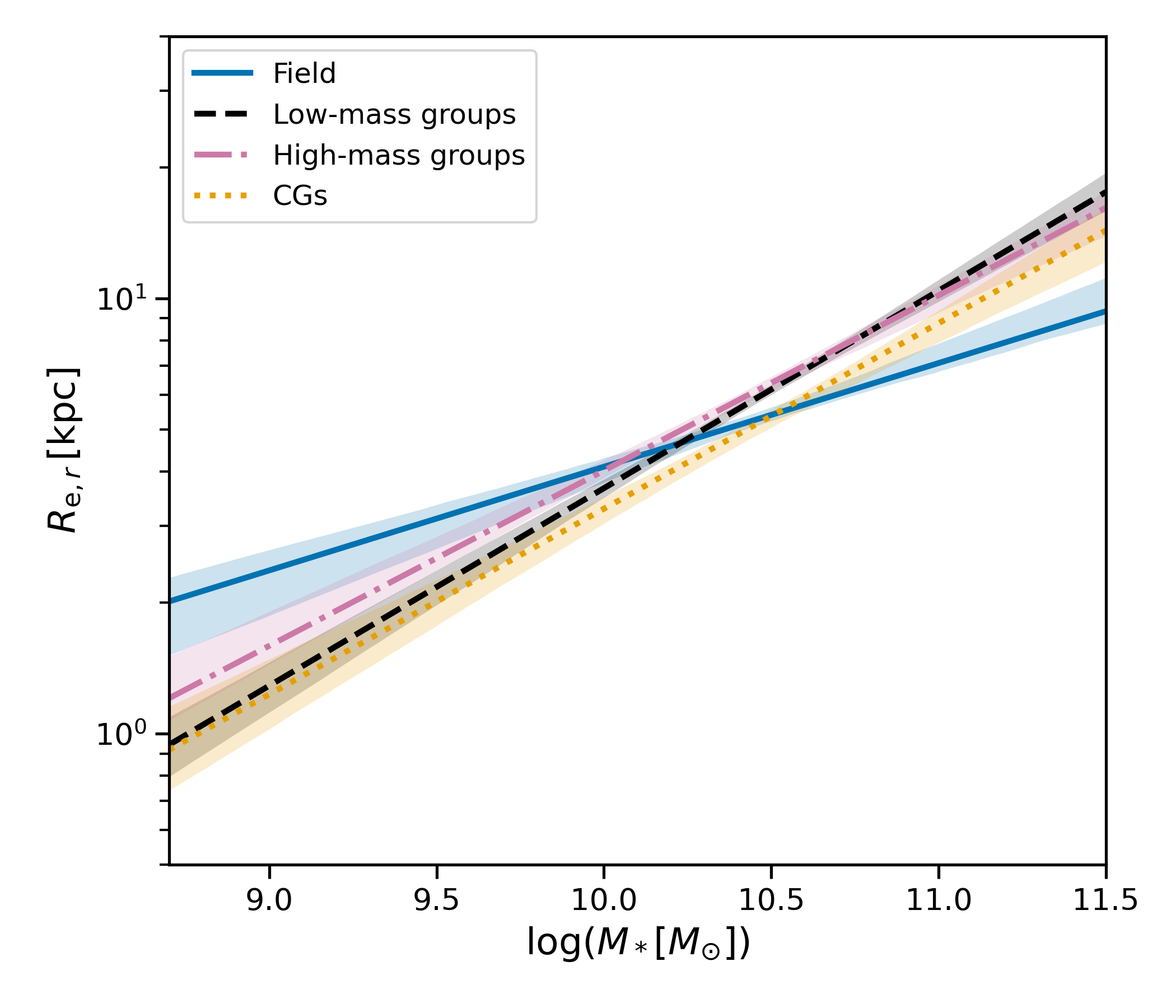} \\
    \includegraphics[width=0.46
    \textwidth]{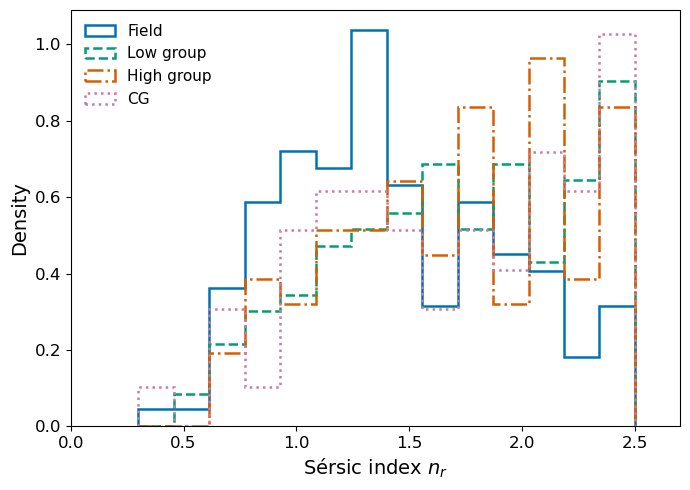}
    \includegraphics[width=0.46
    \textwidth]{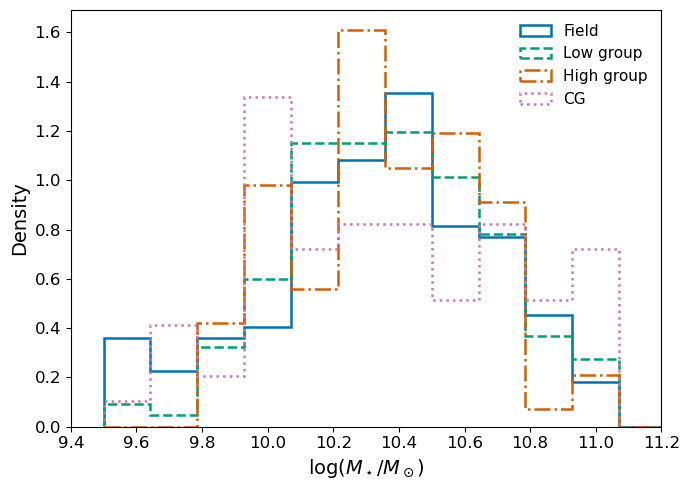}
    \caption{Top panel: Best-fit linear relation between effective radius and stellar mass for TGs in different environments. Bottom-left panel: density distribution of Sérsic index in r-band of TGs for each environment. Bottom-left panel: density distribution of  the logarithm of the stellar mass of TGs.}
    \label{fig:tran_re_m}
\end{figure}

We find no significant differences in the stellar mass–size relation for ETGs, LTGs, and OGs between groups and the field. In contrast, TGs exhibit a clear environmental dependence (see top panel of Fig. \ref{fig:tran_re_m}). In both low-mass and high-mass groups the relation is significantly steeper than in the field, therefore that at low stellar masses TGs in groups are more compact than their field counterparts, whereas at higher masses they become larger at fixed mass. In CGs the slope is also steeper than in the field, while the intercept, i.e. the normalization at the pivot mass $\log(M_\star/M_{\odot}) = 10.5$
, remains consistent with the field within the uncertainties. As a result, the CG and field relations intersect close to $\log(M_\star/M_{\odot}) = 10.5$, TGs in CGs are smaller than field TGs at lower masses and larger at higher masses. These trends indicate that TGs are smaller in denser environments at fixed stellar mass for 
$\log(M_\star/M_{\odot}) \leq 10.5$. However, because the slope is steeper, their sizes grow more rapidly with stellar mass, therefore that difference diminishes and reverses at higher masses.


To further investigate the structural differences of TGs, we analyses the distribution of Sérsic indices across environments. Although we do not perform a full bulge–disc decomposition, the Sérsic index in the r-band provides a useful approximation to the central concentration. The bottom left panel of Fig.~\ref{fig:tran_re_m} shows the $n_r$ distributions for TGs in each environment. In the field, the distribution has a single peak around 
$n_r\sim1.4$, indicating mostly disc-like systems. In both low- and high-mass groups, the distribution shifts to higher $n_r$, starting near $n_r\sim0.5$ and reaching a maximum close to $n_r\sim2$. In compact groups, the distribution is clearly bimodal, with one peak again near $n_r\sim1.4$ and a second peak around $n_r\sim2.2$, consistent with the bimodality reported by \citet{2023Montaguth}. Overall, the $n_r$ distributions of TGs in denser environments are skewed towards higher values compared with the field, indicating that TGs in groups and, especially, in CGs are on average more centrally concentrated, structurally more evolved, and closer to ETGs in their Sérsic indices and mass–size slopes than their field counterparts.

\begin{figure}
    \centering
    \includegraphics[width=0.26\textwidth]{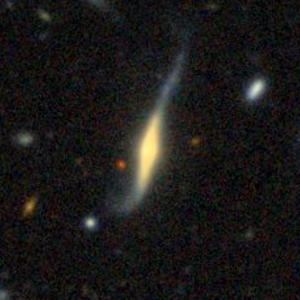}
    \includegraphics[width=0.26\textwidth]{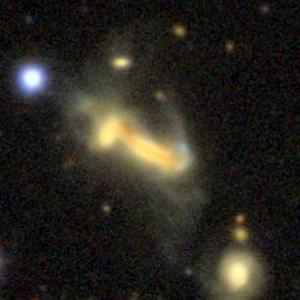}
    \includegraphics[width=0.26\textwidth]{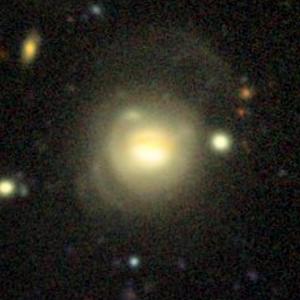}
    \includegraphics[width=0.26\textwidth]{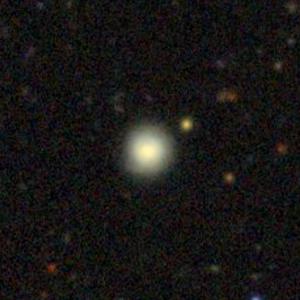}
    \includegraphics[width=0.26\textwidth]{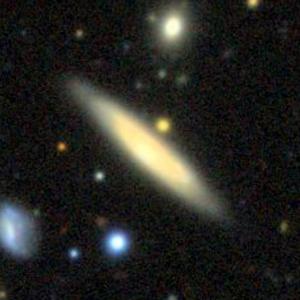}
    \includegraphics[width=0.26\textwidth]{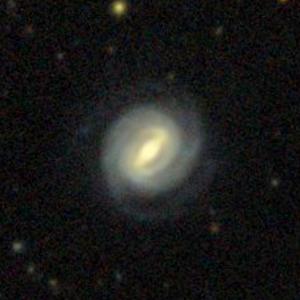}

    \caption{Examples of transition galaxies in the three environments. Each panel shows an Legacy image stamp for a galaxy in a group (left), a CGs (center), and the field (right). The top row presents morphologically disturbed systems. The bottom row shows more regular systems.}
    \label{fig:tran_re_img}
\end{figure}

Another important aspect for understanding this result is the morphological appearance of these galaxies as revealed by inspecting optical images. In the region that we designate as transition galaxies, the sample of \citet{vika2015megamorph} contains only 12 galaxies (5 classified as Sab–Sb, 3 as Sbc–Sc, 2 as S0, and 2 as ellipticals). Although this sample is small, it already suggests that this part of the diagram is populated by a heterogeneous mixture of Hubble types, but the number of objects is far too small to define a robust statistical mapping for our much larger transition class.

The heterogeneous mix of Hubble types in the transition region suggested by the previous work motivates a more detailed morphological analysis here. Therefore, we perform a visual classification of the galaxies in our TG sample using the AstroInspect tool \citep{astroinspect}, which enables the simultaneous inspection of S-PLUS and Legacy Survey images. This inspection shows that, in all environments, our sample is mostly composed of disc-like systems with red colors, with S0-like discs being more common than typical spiral galaxies and with a small contribution of genuinely spheroidal systems. In group and CG environments we further identify a significant number of morphologically perturbed discs, whereas in the field sample, defined following the isolation criterion of \citet{2007yang}, only $\sim 3 \%$ of the galaxies show such merger-like disturbed morphologies; by construction, even these few disturbed galaxies still satisfy the \citet{2007yang} isolation cuts.

In Figure~\ref{fig:tran_re_img} we present six illustrative examples of transition galaxies in the three environments. The top row shows three morphologically disturbed galaxies: from left to right, a galaxy in a group, a disc galaxy in a CG, and a somewhat more spheroidal field galaxy that appears to be the remnant of a major merger, as suggested by its shell-like pattern. Such galaxies are not common in the field, and this object is representative of the $\sim 3\%$ mentioned above. 

To further assess the robustness of the TG classification across environments, and in particular to evaluate the possibility that the TG population in groups could include a contribution from dwarf ellipticals (which typically have $1\leq n\leq2$), we complement the visual examples discussed above with a statistical characterization of the TG population. This analysis allows us to test whether the TG sample could be contaminated by dwarf ellipticals and whether such contamination might vary with environment. In principle, a higher dwarf contribution in groups than in the field could contribute to the environmental differences observed in the TG mass–size relation. However, in our sample, TGs in the field, groups, and CGs span a similar stellar-mass range, $\log(M_\star/M_{\odot})\approx 9.5$–11.0, peaking at $\sim10.3$–10.5 (see the stellar-mass distribution in the bottom-right panel of Fig.~\ref{fig:tran_re_m}), with only a very small fraction of objects in the dwarf-galaxy regime ($\log(M_\star/M_{\odot})\leq 9.45$; \cite{2012McConnachie}). We therefore find no indication that the TG population in groups or CGs is dominated by low-mass systems, and conclude that the observed environmental trends are unlikely to be driven by a varying contribution from dwarf ellipticals.

In addition, the TG population may follow different evolutionary pathways depending on the environment. The lower TG fraction observed in groups and compact groups (CGs) is consistent with a shorter transition timescale in denser environments, where galaxies may evolve more rapidly from the TG phase into the ETG population (e.g., through accelerated quenching accompanied by structural transformation). In this scenario, TGs would be less frequently observed in groups/CGs because the transition phase is shorter-lived, consistent with the accelerated evolution we proposed in \cite{2023Montaguth,2025Montaguth}.

The OG sample also exhibits a mixture of morphologies. Consistently, in the region of the \citet{vika2015megamorph} diagram that most closely overlaps with our OG region, only nine galaxies are found, classified as Sab–Sb (5 objects), Sbc–Sc (3 objects), and S0 (1 object). This confirms that the region we denote as “Other Galaxies” is populated by a heterogeneous set of Hubble types, but in practice it is still largely dominated by late-type disc galaxies.

Following the same visual inspection procedure, we find that this class of galaxies includes regular spiral galaxies, compact blue objects, and early-type systems that still show signs of ongoing star formation. We also identify morphologically perturbed galaxies within the OG class, but they represent a smaller fraction than in the TGs and tend to be bluer. Most OG objects correspond to relatively large, face-on spiral galaxies with ongoing star formation, whose strong central light concentration—produced by bulges or bars—shifts them into the OG region of the Sérsic–colour diagram.

\begin{figure}
    \centering
    \includegraphics[width=0.26\textwidth]{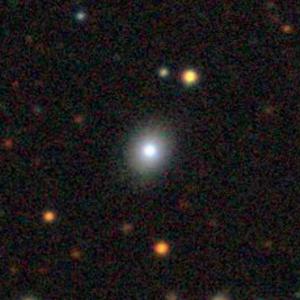}
    \includegraphics[width=0.26\textwidth]{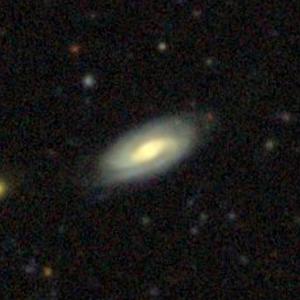}
    \includegraphics[width=0.26\textwidth]{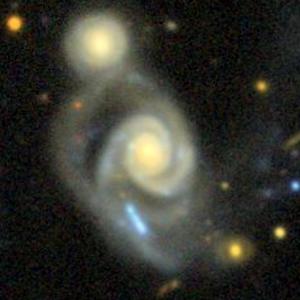}
    \includegraphics[width=0.26\textwidth]{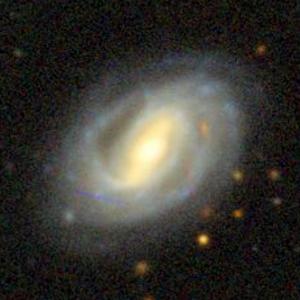}
    \includegraphics[width=0.26\textwidth]{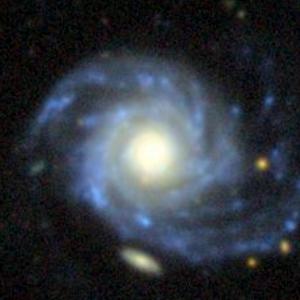}  
     \includegraphics[width=0.26\textwidth]{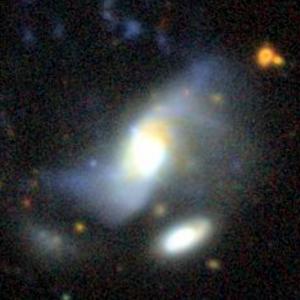}

    \caption{Examples of other galaxies in the three environments. Each panel shows an Legacy image stamp for a galaxy in the field (left), a group (center), a CG (right).} 
    \label{fig:og_re_img}
\end{figure}

These findings support the interpretation of both TG and OG as populations ``in transition'', in the sense that they include galaxies undergoing morphological and star-formation transformation. However, the most advanced stages of this transformation appear within the TG class, which contains a higher fraction of S0-like systems and galaxies with strong visual signatures of interaction. OG galaxies, by contrast, are more strongly dominated by face-on spirals and less frequent, milder disturbances. Representative examples of OG systems in the different environments are shown in Figure~\ref{fig:og_re_img}, illustrating the heterogeneous nature of this class, which is nevertheless predominantly composed of blue, disc-dominated systems, together with a minority of more compact blue galaxies.

\section{Discussion}
\label{sec:discussion}

Our results confirm that the stellar mass–size relation is strongly modulated by environment in the case of TGs. For ETGs, we find no significant differences across environments. This suggests that, by the present epoch, the structural properties of ETGs are primarily governed by stellar mass, with little residual imprint from environment, consistent with previous studies \citep{2013Huertas-company, 2013Huertas-Companyb, 2013Huertas-company}. For LTGs, the lack of strong environmental dependence agrees with \cite{2025Abdullah}. However, for LTGs in compact groups the intercept is lower than in the field, indicating that CG galaxies are smaller at fixed stellar mass. This is consistent with previous results showing that, due to strong galaxy–galaxy interactions, galaxies in compact groups tend to be more compact than their counterparts in the field and in loose groups \citep{2012Coenda,2023Montaguth}. It is important to note that other studies do report environmental trends, for instance, differences between cluster and field LTGs at low stellar masses \citep{2010Maltby}, or more generally across cluster populations \citep{2017Kuchner}. Such discrepancies may arise from differences in how LTGs are defined, or because in dense environments like clusters, LTGs may already be transforming into TGs, thereby imprinting environmental signatures on the mass–size relation. Future analyses applying a homogeneous methodology to both groups and clusters will be essential to clarify these scenarios.

In contrast, TGs exhibit a strong environmental dependence in their mass–size relation. The slope observed for TGs in groups and CGs closely resembles that of ETGs, while TGs in the field follow a slope more similar to LTGs. This suggests that TGs in denser environments may be structurally evolving into ETGs, which could indicate a more rapid transformation from LTGs to ETGs in groups and CGs compared to the field. 

In these group environments, the redder colors of TGs, likely linked to their lower star formation rates compared to LTGs \citep{2023Montaguth}, suggest that they are generally gas-poor systems, whose stellar disks can be more easily altered during gravitational interaction \citep{Moore_1996, 2017Kuchner, 2022lokas, Xu_2025}. Consequently, in groups tidal interactions may play a key role in their evolution, favoring prolonged and repeated encounters that induce structural changes, such as outer stellar stripping or bulge growth through minor mergers \citep{2015Kannan}.

In group-scale environments, particularly low-mass groups and CGs, two distinct processes may be at play in shaping the observed trends. For low-mass galaxies, tidal effects can induce impulsive heating, stellar stripping, and fading of the outer disk, leading to more compact morphologies and smaller sizes. Simultaneously, some TGs may experience minor mergers and gas accretion whose impact depends on the geometry and angular momentum of the accreted material: coplanar mergers and aligned gas inflows tend to grow and thicken the disc \citep[e.g.][]{2008Villalobos}, whereas misaligned or low–angular-momentum accretion, as well as centrally concentrated satellite infall, can drive gas and stars towards the inner regions and enhance central mass concentration \citep[e.g.][]{2003Scannapieco}. Theses processes contribute to the build-up of central stellar mass and likely enhance the prominence of the bulge component—consistent with the observed increase in the fraction of TGs with Sérsic indices $n_r > 2$ in low-mass groups and CGs. These competing mechanisms may help explain the steepening of the mass–size relation in groups, where some galaxies become more compact, while others grow in size.

In high-mass groups, where faster encounters, harassment, and ram-pressure stripping dominate the environmental processes, the mechanisms affecting galaxy evolution differ from those in low-mass groups, but can still contribute to morphological transformation. In this mass regime—especially in systems with halo masses around $10^{13.5}M_{\odot}$—ram-pressure stripping becomes important \citep{2013bahe, 10.1093/mnras/stv100, 2022Kolcu}. This process can remove a large fraction of the gas from the galactic disk, preventing further growth in galaxy size, while leaving the central bulge largely unaffected \citep{2012Steinhauser}. As a result, galaxies may appear more compact and display higher Sérsic indices, placing them within our intermediate-$n$ range. Indeed, \citet{2021Roman-oliveira} find that galaxies affected by ram-pressure stripping, with stellar masses above $10^{9.5}~M_{\odot}$, have a median value of $n = 1.24$. In addition, repeated high-speed tidal encounters (harassment) can heat and partially strip the stellar disc and/or build up the bulge component, effectively increasing the central concentration and Sérsic indices and moving systems into high-$n$ regime \citep{1996Moore,2005Mastropietro}. This suggests that some TGs may have undergone similar environmental processes, which depleted a large fraction of the gas available for star formation and altered their structural properties.

\section{Summary and Conclusions}
\label{sec:summary}

We investigate how the stellar mass–size relation varies across galaxy environments in the local Universe, focusing on compact groups (CGs), low- and high-mass groups, and the field, within the redshift range $0.035 < z < 0.096$. Our sample comprises 595 galaxies in CGs, 1668 in low-mass groups, 834 in high-mass groups, and 915 in the field. Using S-PLUS observations and Sérsic-based morphological classifications, galaxies were divided into early types (ETGs; $n\geq2.5$ and $(u - r)_0 \geq 2.2$), late types (LTGs; $n<2.5$ and $(u - r)_0 < 2.2$), transition galaxies (TGs; $(u - r)_0 \geq 2.2$ and $n<2.5$), and other galaxies (OGs; $(u - r)_0 < 2.2$ and $n\geq2.5$):

\begin{enumerate}
    \item  Our analysis shows that ETGs, and OGs exhibit no significant environmental dependence in their mass–size relations, with consistent slopes and intercepts across environments.
    \item TGs display a strong environmental signature. Their mass–size relation becomes significantly steeper in compact groups and in both low- and high-mass groups compared to the field. As a consequence, TGs in denser environments are smaller at fixed stellar mass below the pivot ($\log(M_*/{\rm M_{\odot}})\leq10.5$), whereas at higher masses the steeper slope implies that their sizes increase more rapidly with stellar mass, thus that the difference with the field diminishes and can reverse for $\log(M_*/{\rm M_{\odot}})\geq10.5$. Our Bayesian regression analysis shows that $\Delta \beta$ excludes zero in compact groups and in both low- and high-mass groups. In contrast, $\Delta \alpha$ is environment-dependent and remains consistent with zero in compact groups. This behaviour is consistent with a scenario in which TGs in dense environments undergo structural evolution, potentially associated with enhanced bulge prominence and outer-disc fading. In line with this interpretation, the Sérsic index distributions show that TGs in groups and compact groups have systematically higher $n_r$ than TGs in the field.

\end{enumerate}

The structural differences observed in TGs relative to field galaxies, first identified in \citep{2023Montaguth} for the specific case of CGs, are here confirmed and generalized. The novelty of this work is to extend the evidence that TGs represent a peculiar and environmentally driven population beyond CGs, establishing the mass–size relation as a key diagnostic of their structural evolution across diverse group contexts. These structural differences likely result from a combination of disk truncation—caused by tidal interactions or ram-pressure stripping—and the preservation or growth of the central bulge, with both processes further enhanced by ongoing interactions and mergers. TGs thus represent a heterogeneous class shaped by multiple environmental mechanisms \citep{2023Montaguth,2025Montaguth,Montaguth2025b}. However, the photometric Sérsic index cannot distinguish classical (merger-built) bulges from pseudo-bulges (disk-evolved), which limits our ability to fully trace their formation pathways. Spatially resolved kinematics from Integral Field Spectroscopy would provide the most direct test, but obtaining statistically large and homogeneous IFS samples for galaxies in diverse group environments is challenging. By contrast, Rubin Observatory LSST will deliver deep, uniform imaging in the near future, allowing us to carry out bulge–disc decompositions and to push structural measurements into the low–surface-brightness regime. This will enable a systematic and scalable assessment of whether the observed mass–size differences are primarily driven by bulge growth, disc stripping, or a combination of both.

\begin{acknowledgements}

We thank the anonymous referee for their constructive comments, which have greatly improved this paper. G.M. gratefully acknowledges the Fundação de Amparo à Pesquisa do Estado de São Paulo (FAPESP) for the support grant 2024/10923-3 and 2025/14602-0. CMdO acknowledges funding through FAPESP 2019/26492-3, 2023/05087-9 and CNPq 307879/2025-9 grants. CL-D acknowledges support from the Agencia Nacional de Investigación y Desarrollo (ANID) through Fondecyt project 3250511. STF acknowledges the financial support of DIDULS/ULS through a regular project number PR2453858 and the funding ADI2553855. AM acknowledges support from the ANID FONDECYT Regular grant 1251882,  from the ANID BASAL project FB210003, and funding from the HORIZON-MSCA-2021-SE-01 Research and Innovation Programme under the Marie Sklodowska-Curie grant agreement number 101086388. F.R.H. acknowledges support from FAPESP grants 2018/21661-9 and 2021/11345-5. Y.J-T. acknowledges financial support from the State Agency for Research of the Spanish MCIU through Center of Excellence Severo Ochoa award to the Instituto de Astrofísica de Andalucía CEX2021-001131-S 
funded by MCIN/AEI/10.13039/501100011033, and from the grant PID2022-136598NB-C32 Estallidos and project ref. AST22-00001-Subp-15 funded by the EU-NextGenerationEU.

We are grateful to Cheng Cheng, Alvaro Alvarez-Candal, Ricardo Demarco, Swayamtrupta Panda, Amanda Lopes, and Marco Grossi for their insightful comments and suggestions, which significantly helped to improve this paper. 

The S-PLUS project, including the T80-South robotic telescope and the S-PLUS scientific survey, was founded as a partnership between the Fundação de Amparo à Pesquisa do Estado de São Paulo (FAPESP), the Observatório Nacional (ON), the Federal University of Sergipe (UFS), and the Federal University of Santa Catarina (UFSC), with important financial and practical contributions from other collaborating institutes in Brazil, Chile (Universidad de La Serena), and Spain (Centro de Estudios de Física del Cosmos de Aragón, CEFCA). We further acknowledge financial support from the São Paulo Research Foundation (FAPESP), the Brazilian National Research Council (CNPq), the Coordination for the Improvement of Higher Education Personnel (CAPES), the Carlos Chagas Filho Rio de Janeiro State Research Foundation (FAPERJ), and the Brazilian Innovation Agency (FINEP).

The authors are grateful for the contributions of CTIO staff in helping in the construction, commissioning and maintenance of the T80-South telescope and camera. We are also indebted to Rene Laporte and INPE, as well as Keith Taylor, for their important contributions to the project. We also thank CEFCA staff for their help with T80-South. Specifically, we thank Antonio Marín-Franch for his invaluable contributions in the early phases of the project, David Cristóbal-Hornillos and his team for their help with the installation of the data reduction package jype version 0.9.9, César Íñiguez for providing 2D measurements of the filter transmissions, and all other staff members for their support.
\end{acknowledgements}

%
%

\bibliography{sample701}{}
\bibliographystyle{aasjournal}



\end{document}